\documentclass[submitting]{nst}

\usepackage{subfigure,dcolumn}
\usepackage[T2A,T1]{fontenc}
\usepackage[russian,english]{babel}
\usepackage{makecell}

\usepackage{epsf}

\usepackage{listings}
\begin{document}

\title{{Design of a 200-MHz Continuous-Wave Radio Frequency Quadrupole Accelerator for Boron Neutron Capture Therapy}}\thanks{Supported by the National Natural Science Foundation of China (No. 11535016, No. 11675236, No. 12075296 and No.11775284)}

\author{Zhi-chao Gao}
\affiliation{Institute of Modern Physics, Chinese Academy of Science, Lanzhou 730000, China}
\affiliation{University of Chinese Academy of Sciences, Beijing 100049, China}

\author{Liang Lu}
\email[Corresponding author, ]{luliang3@mail.sysu.edu.cn}
\affiliation{The Sino-French Institute for Nuclear Energy and Technology, Sun Yat-sen University, Zhuhai 519000, China}

\author{Chao-chao Xing}
\affiliation{Institute of Modern Physics, Chinese Academy of Science, Lanzhou 730000, China}%
\author{Lei Yang}
\affiliation{Institute of Modern Physics, Chinese Academy of Science, Lanzhou 730000, China}%
\author{Tao He}
\affiliation{Institute of Modern Physics, Chinese Academy of Science, Lanzhou 730000, China}%
\author{Xue-ying Zhang}
\affiliation{Institute of Modern Physics, Chinese Academy of Science, Lanzhou 730000, China}\affiliation{University of Chinese Academy of Sciences, Beijing 100049, China}

\begin{abstract}
	A high-intensity continuous wave (CW) radio frequency quadrupole (RFQ) accelerator is designed for boron neutron capture therapy (BNCT). The transmission efficiency of a 20-mA proton beam accelerated from 30 keV to 2.5 MeV can reach 98.7\% at an operating frequency of 200 MHz. The beam dynamics has good tolerance to errors. By comparing the high-frequency parameters of quadrilateral and octagonal RFQ cross-sections, the quadrilateral structure of the four-vane cavity is selected owing to its multiple advantages, such as a smaller cross section at the same frequency and easy processing. In addition, tuners and undercuts are designed to tune the frequency of the cavity and achieve a flat electric field distribution along the cavity. In this paper, the beam dynamics simulation and electromagnetic design are presented in detail.
\end{abstract}


\maketitle

\nolinenumbers
\section{Introduction}\label{sec:Introduction}

	As a cancer treatment method, boron neutron capture therapy (BNCT), which is a bimodal form of radiation therapy was first proposed by the American scientist Locher in 1936~\cite{locher1936biological}. Because this approach requires drugs containing $\rm ^{10}B$ and an epithermal neutron beam that can kill tumor cells without affecting other tissues, a safe and stable neutron source is necessary. 
	
	Early BNCT experiments and clinical applications have used reactors as neutron sources, such as the Massachusetts Institute of Technology Reactor II ~\cite{Rogus1994Mixed} and the Tehran Research Reactor ~\cite{KASESAZ2014132}. In recent years, with the development of strong current accelerator technology, BNCT neutron sources based on accelerators have been rapidly developed. Researchers have proposed a method for using an RFQ accelerator~\cite{1969A} to accelerate proton beams bombarding on targets to produce neutrons, which can be used as the front-end accelerator of AB-BNCT neutron source facilities ~\cite{2002Conceptual}.For
example, INFN (Istituto Nazionale di Fisica Nucleare) in
Italy developed the TRASCO RFQ (TRAsmutazione SCOrie
RFQ) for BNCT research~\cite{2004TRASCO}. INMRC (Ibaraki Neutron Med-
ical Research Center) in Japan constructed a strong current
proton accelerator which consists of a RFQ and a DTL for
the same purpose~\cite{2014Japan}. These low-cost accelerator-based (AB) neutron sources with a compact structure are more suitable for use in hospitals.
	
	A four-vane RFQ accelerator with an operating frequency of 200 MHz was designed in this study. The dynamics properties are presented in detail in the following section. In section~\ref{sec:Electromagnetic design}, we describe the electromagnetic design, including the cross-section, tuners, and undercuts.

	\section{Beam dynamics design}\label{sec:Beam dynamics design}
	
	\subsection{Requirements of BNCT}\label{subsec:Requirements of BNCT}
	Beryllium or lithium is chosen as the target marterial of
AB-BNCT neutron source in most cases. Although beryllium targets achieve a better thermal performance and mechanical properties, we chose to use the lithium target because it requires a lower output energy for the accelerator. The reaction $\rm ^7Li(p,n)^7Be$ was selected to produce neutrons for BNCT because of its high neutron yield and relatively soft neutron energy spectrum, which can be used in BNCT clinical treatment with no significant slowing down. More considerations are involved in the design of lithium targets, which are not the focus of this paper and are not be discussed here.
	
	According to the requirements of the International Atomic Energy Agency (IAEA) for clinical BNCT, the desirable minimum flux of epithermal neutrons (0.5 eV to 10 keV) is $\rm 10^9 n/cm^2/s$~\cite{IAEA}. The $\rm ^7Li(p,n)^7Be$ reaction has a threshold energy of 1.88 MeV and a remarkable resonance peak at 2.25 MeV for proton. It has a relatively low output energy for the RFQ, which makes it easier to design an RFQ and reduce the cost.
	
	Considering the requirements of the IAEA and the characteristics of the reaction, Peking University has conducted a simulation showing that a 2.5 MeV proton beam with a current of 15 mA can fully meet the requirements for cancer treatment~\cite{aJEON2020633, ZHU201857}. Thus, a beam current of 20 mA was chosen with a final energy of 2.5 MeV.
	
	\begin{table}[htbp] 
		\centering
		\caption{Main parameters of RFQ.} 
		\label{Main parameters of RFQ.}
			\begin{tabular}{ll}
			\toprule
				Parameters              & Value             \\ 
				\midrule
				Particle                &$\rm H^+$             \\
				Frequency (MHz)          &200                \\
				Input energy (keV)       &30                 \\
				Output energy (MeV)      &2.5                \\
				Current (mA)             &20                 \\
				Inter-vane voltage (kV)  &91.65   \\
				Average aperture (mm)    &5.406              \\
				Rv/R0                   &0.75               \\
				Vane length (mm)         &3162.62            \\
				Peak surface electric field (MV/m)& 23.3774 (Kp = 1.59)       \\
				Operation mode          &CW                 \\
				Transmission efficiency &98.7\%             \\
				\bottomrule
			\end{tabular}
	\end{table}

	\begin{figure*}[htbp]
		\centering
		\includegraphics[width=.8\textwidth]{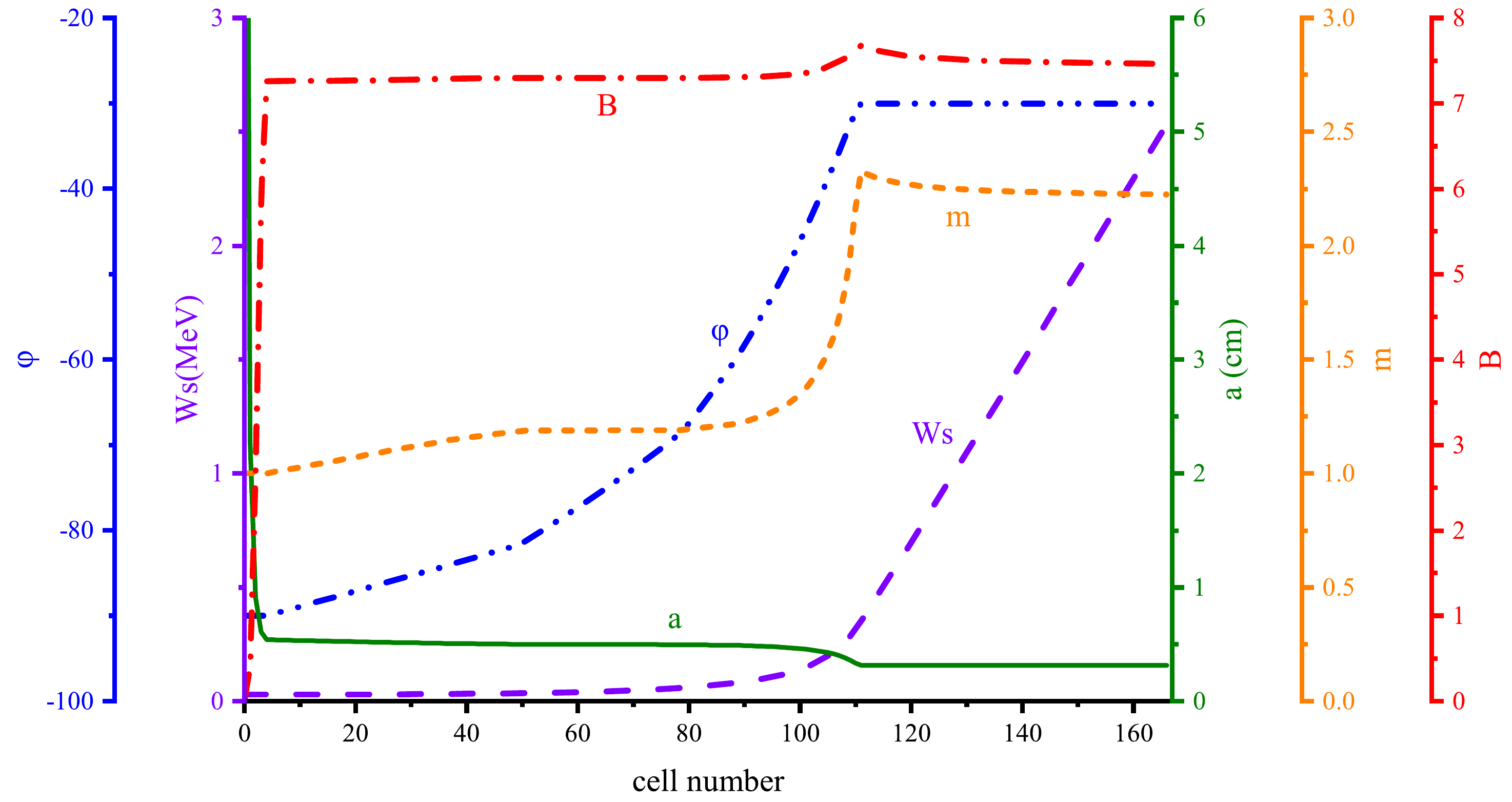}
		\caption{RFQ beam dynamics main parameters vary with accelerating cells.}
		\label{dynamics_parameters}
	\end{figure*}
	
	\subsection{Beam dynamics parameters}\label{subsec:Beam dynamics parameters}
	The main parameters of the BNCT-RFQ are listed in Table~\ref{Main parameters of RFQ.}. As indicated in Chapter ~\ref{subsec:Requirements of BNCT}, the particles to be accelerated, as well as the beam current and output energy, were determined. In addition, the frequency choice directly affects the size of the RFQ cavity, which is related to the cost of the construction. To reduce the occupied area and the construction cost of the RFQ to suit a hospital environment, the frequency is decided to be relatively higher, i.e., 200 MHz. The inter-vane voltage, which is related to the beam transmission efficiency and focusing effect, was determined to be 91.65 kV. Although the inter-vane voltage is relatively high, this parameter is considered reasonable. There are RFQs operating at higher inter-vane voltages. The LEDA RFQ in LANL has an inter-vane voltage ranging from 67 to 117 kV~\cite{1997Simulations}. The IFMIF-EVEDA RFQ in LNL has a minimum voltage of 79 kV and a maximum voltage of 132 kV~\cite{Comunian2008THE}. Meanwhile, the voltage ramp of the FRIB RFQ at Michigan State University ranges from 60 to 112 kV~\cite{2015DESIGN}. Portions of their operating inter-vane voltages are higher than 90 kV, which indicates that our design is reliable. A lower vane-gap voltage can improve the stability of the RFQ operation, but the length of the cavity, construction cost, and occupied area will increase. The other parameters listed in the table were determined through dynamics simulation.
	
	\subsection{Beam dynamics design}\label{subsec:Dynamics transmission simulation}
\begin{figure*}[htbp]
		\centering
		\includegraphics[width=.6\textwidth]{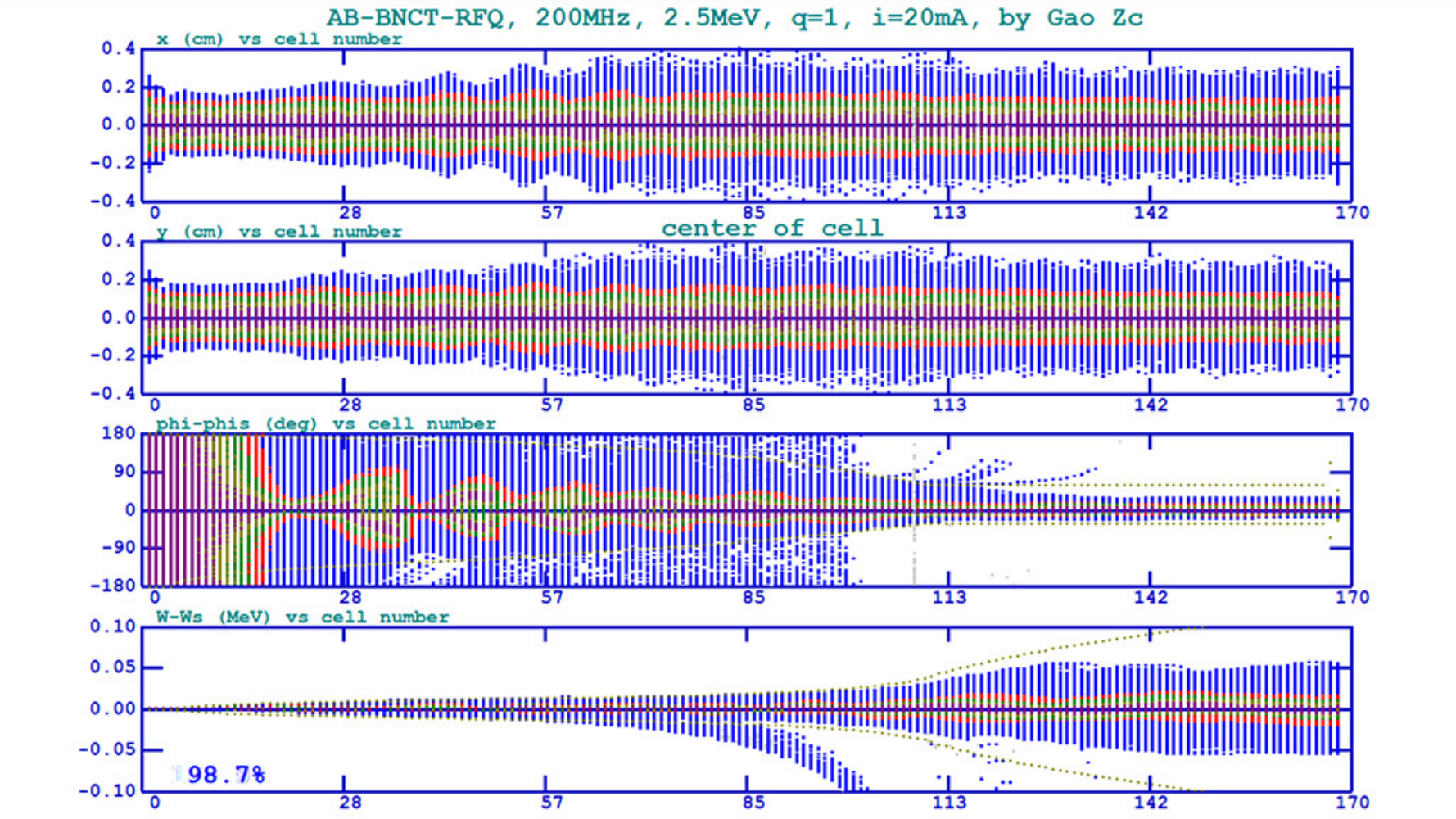}
		\caption{RFQ beam transmission process. There are beam envelope of the X axis and the Y axis, the difference between particle phase and synchronous phase and energy spread, respectively.}
		\label{dynamics1}
	\end{figure*}

\begin{figure*}[htbp]
		\centering
		\includegraphics[width=.6\textwidth]{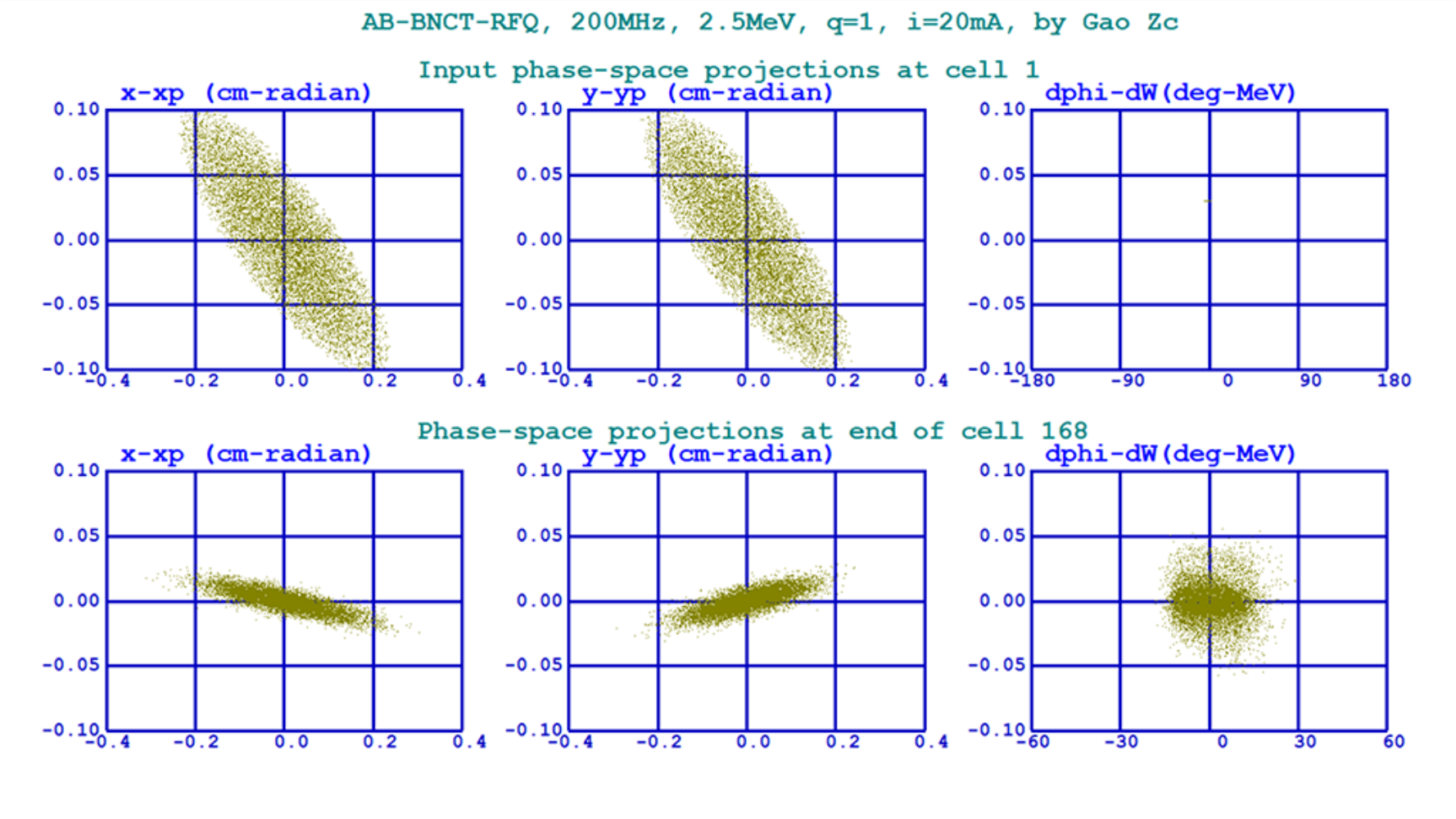}
		\caption{Beam profiles at the entrance and exit of the RFQ.}
		\label{dynamics2}
	\end{figure*}

\begin{figure*}[htbp]
		\centering
		\includegraphics[width=.6\textwidth]{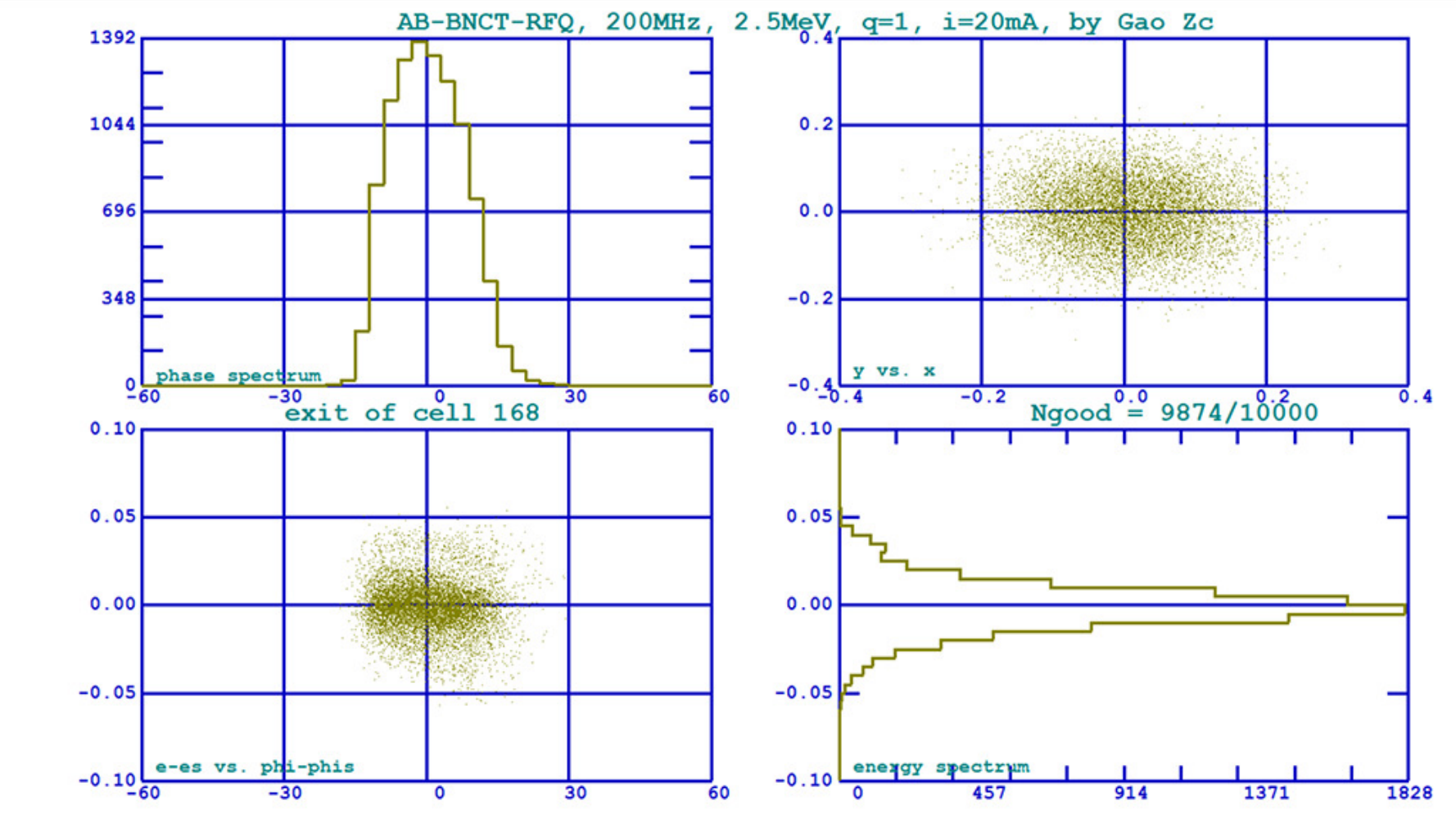}
		\caption{Beam distribution at the exit of the RFQ.}
		\label{dynamics3}
	\end{figure*}

	In this study, a dynamics transmission simulation was conducted using the software RFQGen~\cite{RFQCodes}. Los Alamos National Laboratory (LANL) developed RFQGen based on the four-stage theory proposed by Stokes and Crandall ~\cite{2007The}, which allows the design of accelerating cells of the RFQ, including radial matching sections at both ends of the structure. The code can adjust the two-term potential functions by changing the vane geometry, thus generating both the accelerating force and the focusing force. Fig.~\ref{dynamics_parameters} shows the main parameters of the designed beam dynamics. 
	When considering the shortest length and high transmission efficiency as the design goals, the length of this RFQ (acceleration proton beam of up to 2.5 MeV with an operating frequency of 200 MHz) is 3.2 m. Other similar RFQ accelerators are much longer, including the ADS-RFQ~\cite{2012DESIGN} in the Institute of Modern Physics in China (which operates at a frequency of 162.5 MHz and accelerates the proton beam from 35 keV to 2.1 MeV), which is 4.2 m, and the BNCT-RFQ~\cite{ZHU201857} at Peking University (which operates at 162.5 MHz and provides acceleration of a 20-mA proton beam at up to 2.5 MeV), which is 5.2 m.
	The following figures show the simulation results for the beam dynamics. In Fig.~\ref{dynamics1}, the transmission efficiency reaches 98.7\% when 10000 particles are simulated. In addition, Fig.~\ref{dynamics2} and Fig.~\ref{dynamics3} shows the phase-space distribution and beam distribution at the entrance and exit of the RFQ.

	\subsection{Dynamics design tolerance}\label{subsec:Redundancy of dynamics}
	\begin{figure*}[htbp]
		\centering
		\includegraphics[width=0.8\textwidth]{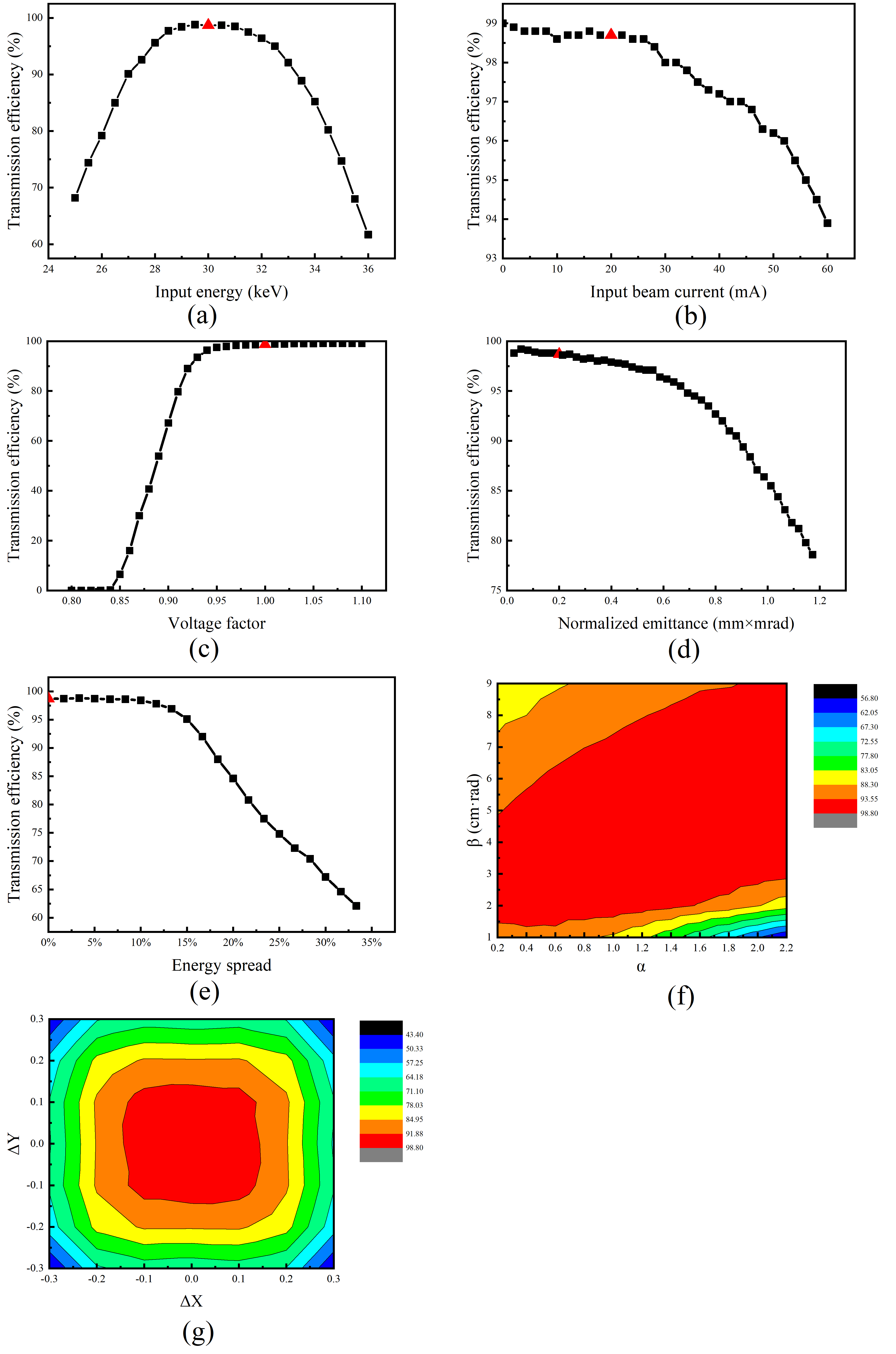}
		\caption{The effects of the (a) input energy, (b) input beam current, (c) normalized vane voltage, (d) normalized emittance, (e) energy spread, (f) Twiss parameters ($\alpha,\beta$), and (g) spatial displacements on the transmission efficiency.}
		\label{seven figures}
	\end{figure*}
	
	Only the ideal states of the parameters are considered in the dynamics transmission simulation. However, a slight disturbance will occur with the parameters when the beam travels from the ion source and low-energy beam transport line (LEBT) to the entrance of the RFQ. This may affect the transmission efficiency of the RFQ; thus, it is necessary to study whether the effect of the non-ideal matching conditions on the beam transmission efficiency is acceptable. Seven parameters (disturbance of input beam energy, current, voltage factor, normalized emittance, energy spread, Twiss parameters, and spatial displacements) were considered in this study. Fig.~\ref{seven figures} shows the results (the red triangles represent the designed parameters).

	The input beam energy has an impact on the synchronous phase. Therefore, a deviation of the input energy may change the synchronous conditions, which leads to beam loss. As shown in Fig.~\ref{seven figures}(a), the beam transmission efficiency is below 90\% when the input energy is greater than 33 keV or less than 27 keV. The beam current is related to the space-charge effects, which increase with the beam intensity. As shown in Fig.~\ref{seven figures}(b), the beam transmission efficiency is over 98\% as long as the beam current is lower than 30 mA. Fig.~\ref{seven figures}(c) shows the change in beam transmission efficiency when the normalized voltage factor varies from 0.8 to 1.1, where a voltage factor of 1.00 indicates 91.65 kV. The beam transmission efficiency is acceptable if the vane voltage is 0.93-times higher than the designed voltage. Fig.~\ref{seven figures}(d) shows the relationship between the normalized emittance of the beam and beam transmission efficiency. To ensure that the beam transmission efficiency is greater than 90\%, the transverse emittance of the input beam should be no greater than 0.9 mm$\cdot$mrad. In addition, the beam transmission efficiency has sufficient margins for the beam energy spread, Twiss parameters, and beam spatial displacement, as shown in Figs.~\ref{seven figures}(e)–(g). Overall, the parameters in Fig.~\ref{seven figures} have a good margin for the beam transmission efficiency, which can guarantee a neutron flux for BNCT treatment.
	
	\section{Selection of cavity}\label{sec:Selection of cavity}
	Currently, the main cavity types of RFQ are four-vane RFQ~\cite{2015Japan,aSNSRFQ,LI201738} and four-rod RFQ~\cite{CHAKRABARTI2004599,2011Simulation,IHRFQ}. The four-vane RFQ is suitable for accelerating light particles, such as protons and deuterons, and can operate under continuous-wave (CW) conditions. Compared with the four-rod RFQ, the four-vane RFQ has the advantages of a higher mechanical strength, lower power loss and power density, and simpler water-cooling structure. At the same frequency, the Q value of the four-vane RFQ cavity is larger~\cite{ma2017design}. However, the four-vane structure has certain disadvantages, such as a large volume and high cost at low frequencies. For the AB-BNCT neutron source, the accelerator needs to operate in CW mode, considering that the cavity radius is approximately 15 cm after a preliminary simulation for a design frequency of 200 MHz, which is acceptable, the four-vane RFQ is selected as the injector for the BNCT neutron source.
	
	\subsection{Quadrilateral cavity}\label{subsec:Quadrilateral cavity}
\begin{figure*}[htbp]
		\centering
		\subfigure[]{
			\includegraphics[width=0.45\textwidth]{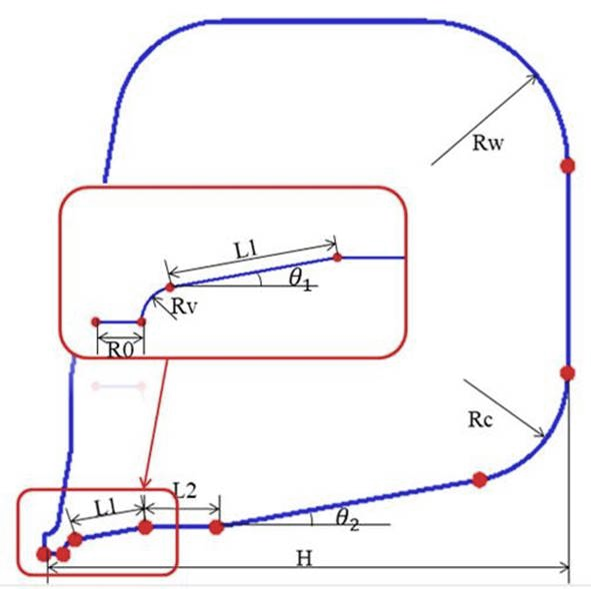}
			\label{quadrilateral_RFQ}
		}
		\subfigure[]{
			\includegraphics[width=0.45\textwidth]{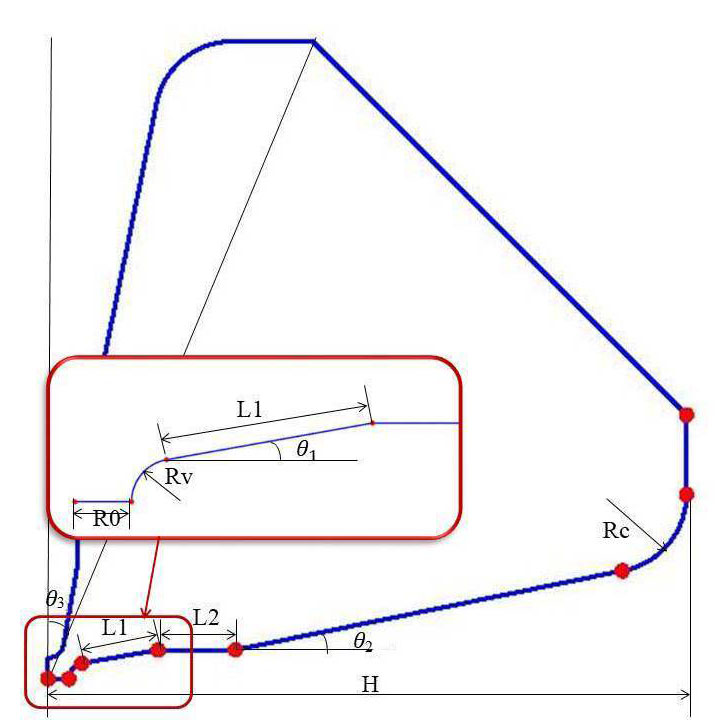}
			\label{octagonal_RFQ}
		}
		\caption{Cross-sections of (a) quadrilateral RFQ and (b) octagonal RFQ.}
		\label{cross-sections}
	\end{figure*}

	A quadrilateral four-vane RFQ has a cross section similar to a square. Its right angles are replaced by arc angles to avoid point discharge and local over-temperature. Because the cross section is centrosymmetric, a quarter of the sketch is shown in Fig.~\ref{quadrilateral_RFQ}, which is determined using nine parameters. The values of the parameters determine the shape of the cross-section. The values are presented in Table~\ref{The cross-section parameters of quadrilateral RFQ and octagonal RFQ.}.

	\begin{table}[htbp]
		\centering 
		\caption{Cross-section parameters of quadrilateral RFQ and octagonal RFQ.} 
		\label{The cross-section parameters of quadrilateral RFQ and octagonal RFQ.}
			\begin{tabular}{lll} 
			\toprule
				Parameter  & Quadrilateral RFQ & Octagonal RFQ  \\
				\midrule
				R0 (mm)           & 5.406    & 5.406  \\
				Rv (=0.75*R0) (mm) & 4.05    & 4.05   \\
				L1 (mm)           & 20    & 20     \\
				L2 (mm)          & 20    & 20     \\
				$\theta_1$ ($^\circ$)       &10       & 10   \\ 
				$\theta_2$ ($^\circ$)      &10        &10      \\ 
				$\theta_3$ ($^\circ$)     &null   &22.5    \\
				Rc (mm)          & 30    &30      \\ 
				Rw (mm)          & 40    &null      \\
				H (mm)           & 147.295    &164.8 \\
			\bottomrule
			\end{tabular} 
	\end{table}

	\subsection{Octagonal cavity}\label{subsec:Octagonal cavity}
	The octagonal four-vane RFQ is an RFQ with a cross section similar to an octagon. From the sketch of the cross-section, it is similar to the quadrilateral cavity except that the arc angles at the four peaks are replaced by straight lines, as shown in Fig.~\ref{octagonal_RFQ}). Table~\ref{The cross-section parameters of quadrilateral RFQ and octagonal RFQ.} lists the design parameters, which are similar to those of the quadrilateral cavity. Here, $\theta_3$ is designed to have a larger value to reduce the power loss and local power density.

	\subsection{Comparison}\label{subsec:Comparison}
	To determine which type of cavity is better in terms of performance, a comparison of high-frequency parameters between the quadrilateral RFQ and octagonal RFQ is presented in Table~\ref{Comparison of high frequency parameters between quadrilateral and octagonal cavity.}, where $\Delta f$ indicates the difference in frequency between the nearest quadrupole and dipole modes. The vane is made of copper and has an electrical conductivity of $5.8\times 10^7$ S/m. To reduce the simulation time, we use the performance of a slice of the cavity instead of the entire RFQ cavity. For a more realistic simulation, the boundaries of the cross profiles (lengthwise direction) were set as magnetic boundaries. The similar structure parameters, cavity frequency, and slice thickness of both types were designed in the same way. It is clear that the values of $\Delta f$ are both above 6 MHz. It is likely that neither requires a frequency-separation structure. This is proved in the following section. From the perspective of the cavity size, H of the octagonal slice is 11.4\% larger than that of the quadrilateral slice, which means that the cross-sectional area of the octagonal cavity is 13\% larger than that of the quadrilateral cavity. A smaller size can reduce the costs of the construction and space. In addition, the quadrilateral slice has a higher Q value and less total RF loss. The cooling system for the quadrilateral cavity is easier to design and construct. Considering that the injector will be operated in CW mode, the quadrilateral RFQ cavity is a better choice.
	\begin{table}[htbp]
		\centering
		\caption{Comparison of high-frequency parameters between quadrilateral and octagonal cavity.}
		\label{Comparison of high frequency parameters between quadrilateral and octagonal cavity.}
			\begin{tabular}{lll}
			\toprule
				Parameter     &Quadrilateral RFQ  silce   &Octagonal RFQ slice\\
				\midrule
				Frequency (MHz)    & 200      & 200       \\
				Thickness (mm)     & 5        & 5         \\
				$\Delta f$ (MHz)         & 6.36     & 6.35      \\
				H (mm)             &147.295   & 164.8      \\
				Q                  &15715     & 14782      \\
				\makecell[l]{Total loss normalized\\to whole cavity (kW)}    &79.93     & 85.21     \\
				\bottomrule
			\end{tabular}
	\end{table}
	
	\section{Electromagnetic design}\label{sec:Electromagnetic design}
	
			\begin{figure*}[htbp]
		\centering
		\subfigure[]{
			\includegraphics[width=0.45\textwidth]{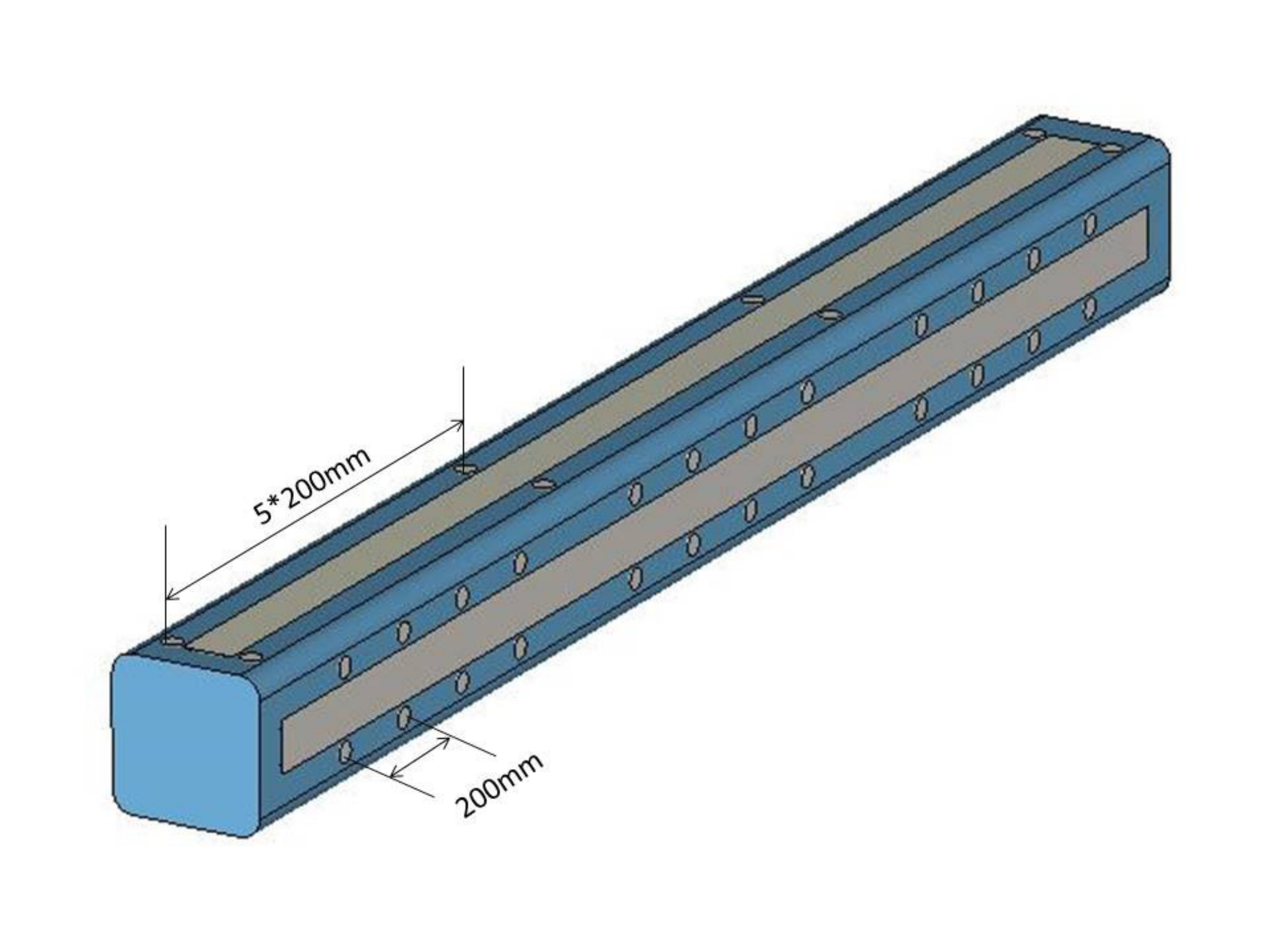}
		}
		\subfigure[]{
			\includegraphics[width=0.45\textwidth]{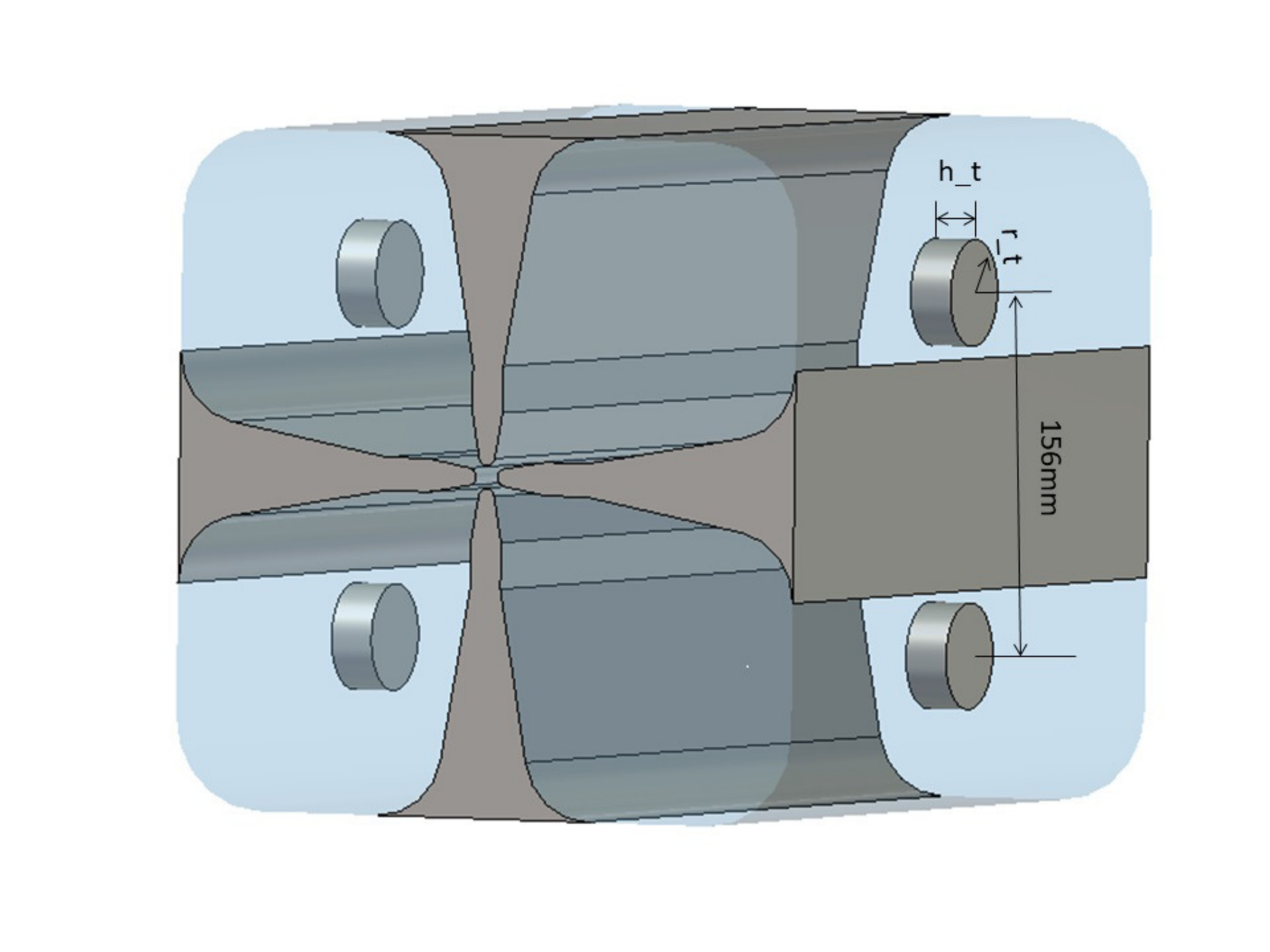}
		}
		\caption{Design diagram of (a) whole cavity and (b) tuners.}
		\label{The design diagram of tuners.}
	\end{figure*}

	The goal of the EM design of the RFQ is to ensure that the frequency of the cavity meets the design frequency by adjusting the parameters or adding tuners and undercuts. It is also important to ensure that the electric field along the beam direction between the two opposite electrodes is flat. The frequency of the cavity and the spatial distribution of the eigenstates of the electric field are related to the shape of the cavity, that is, parameter H. The radio-frequency (RF) structure design of the cavity is to elect the necessary eigenstate by finding out the spatial distribution of the field suitable for accelerating ion beam, and then optimizing the size of the cavity to ensure that the cavity frequency meets the design frequency. The 3D code CST Microwave Studio~\cite{CST} was used for the RF structure design of the RFQ. The parameters shown in Fig.~\ref{quadrilateral_RFQ} and Table~\ref{The cross-section parameters of quadrilateral RFQ and octagonal RFQ.} can be adjusted to optimize the high-frequency performance using MWS software.

	\subsection{tuner design}\label{subsec:tuner design}

	The cylindrical structures shown in Fig.~\ref{The design diagram of tuners.} are the tuners of the RFQ, and Table~\ref{Parameters of tuners and undercuts.} shows the design parameters of the tuners. Tuners are used to slightly tune the overall resonant frequency of the cavity to keep the operating frequency of the cavity stable at the design frequency, or to adjust the local resonant frequency or electric field distribution of the cavity to achieve flatness of the electric field distribution along the cavity.
	
\begin{figure}[htbp]
		\centering
		\includegraphics[width=.48\textwidth]{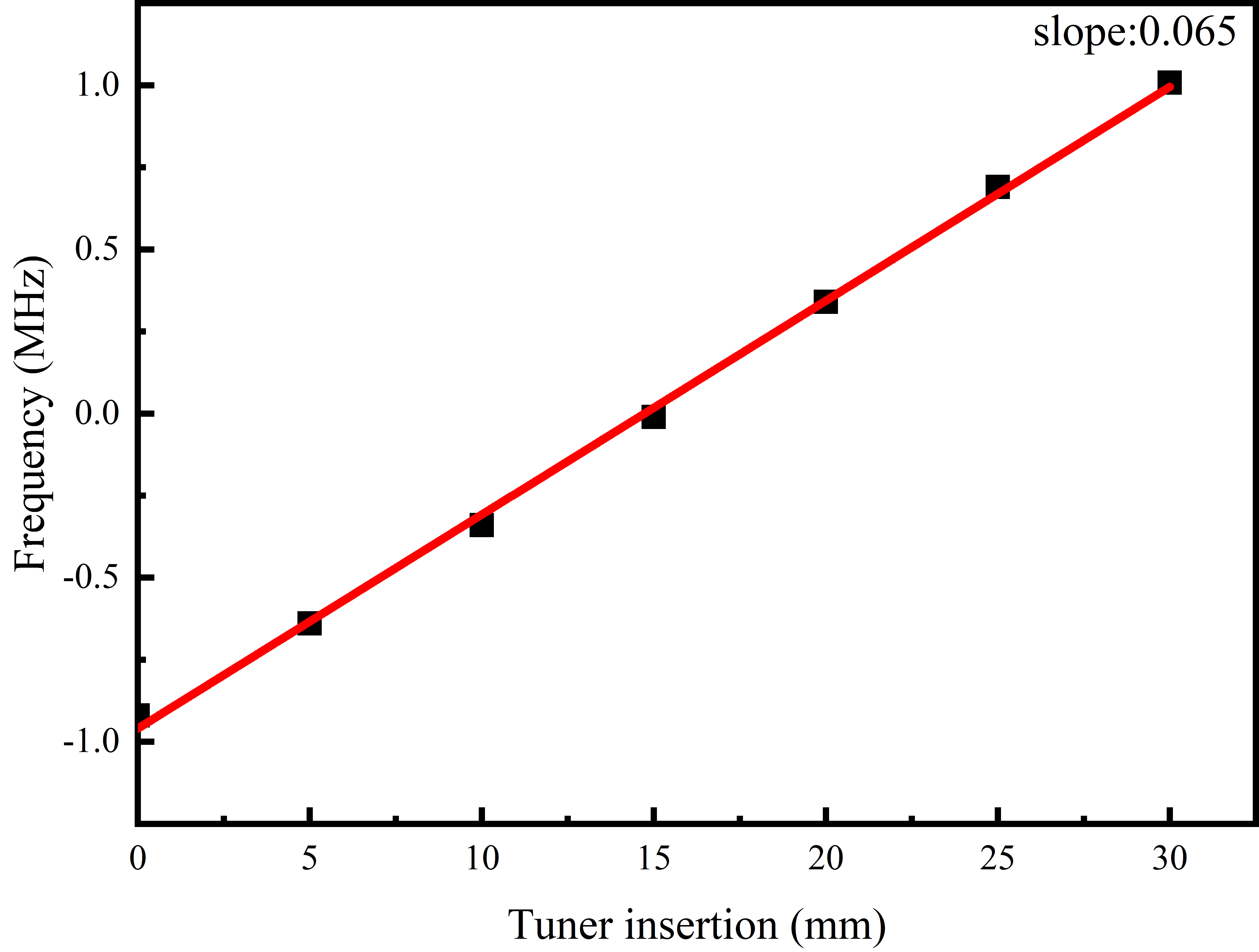}
		\caption{Effect of tuner insertion depth on cavity frequency (for all tuners).}
		\label{Effect of tuner insertion depth on cavity frequency.}
	\end{figure}

	As shown in Fig.~\ref{The design diagram of tuners.}, the RFQ cavity has 64 tuners with a radius of 20 mm and a pre-insertion depth of 15 mm, uniformly distributed over the symmetrical electrodes. The effect of the tuners on the cavity frequency can be studied by adjusting the insertion depth of the tuners. According to Fig.~\ref{Effect of tuner insertion depth on cavity frequency.}, there is a linear relationship between the insertion depth and the cavity frequency, and the average sensitivity is 1 kHz/mm per tuner, that is, the tuning range of the whole cavity is -0.92 to 1.0075 MHz.
	
	As indicated in section ~\ref{subsec:Comparison}, there is no need for a frequency-separation structure. This is proven in chapter \ref{subsec:The whole cavity simulation} after the whole cavity simulation has been completed.

	\begin{table}[htbp]
		\centering 
		\caption{Parameters of tuners and undercuts.} 
		\label{Parameters of tuners and undercuts.}
			\begin{tabular}{lll} 
			\toprule
				~            & Parameter  & Value   \\
				\midrule
				Tuner        &r\_t (mm)           & 20  \\
				~            &h\_t (mm)           & 15  \\
				~            &Longitudinal interval between adjacent tuners (mm)      & 200     \\
				~            &Tuner intervals in adjacent quadrants (mm)            & 156     \\ 
				Undercut     &h\_cut (mm)         &60    \\
				~            &theta ($^{\circ}$)  &60    \\
				~            &D\_in (mm)          &75    \\
				~            &D\_out (mm)         &68    \\
				\bottomrule
			\end{tabular} 
	\end{table}

	\subsection{Undercut design}\label{subsec:Undercut design}
	\begin{figure*}[htbp]
		\centering
		\includegraphics[width=0.8\textwidth]{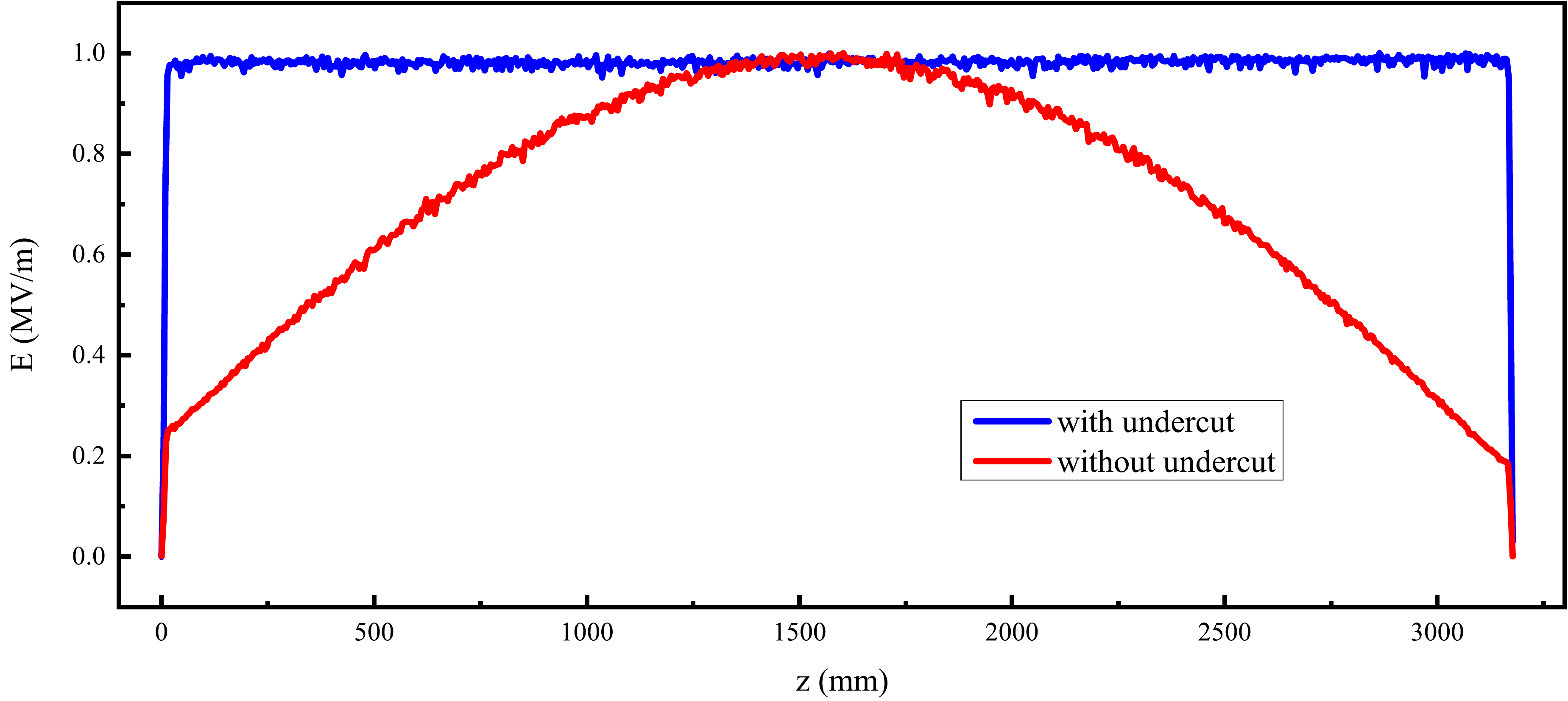}
		\caption{Normalized electric field distribution between electrodes with and without undercut.}
		\label{undercut12}
	\end{figure*}
	
	\begin{figure}[htbp]
		\centering
		\includegraphics[width=0.5\textwidth]{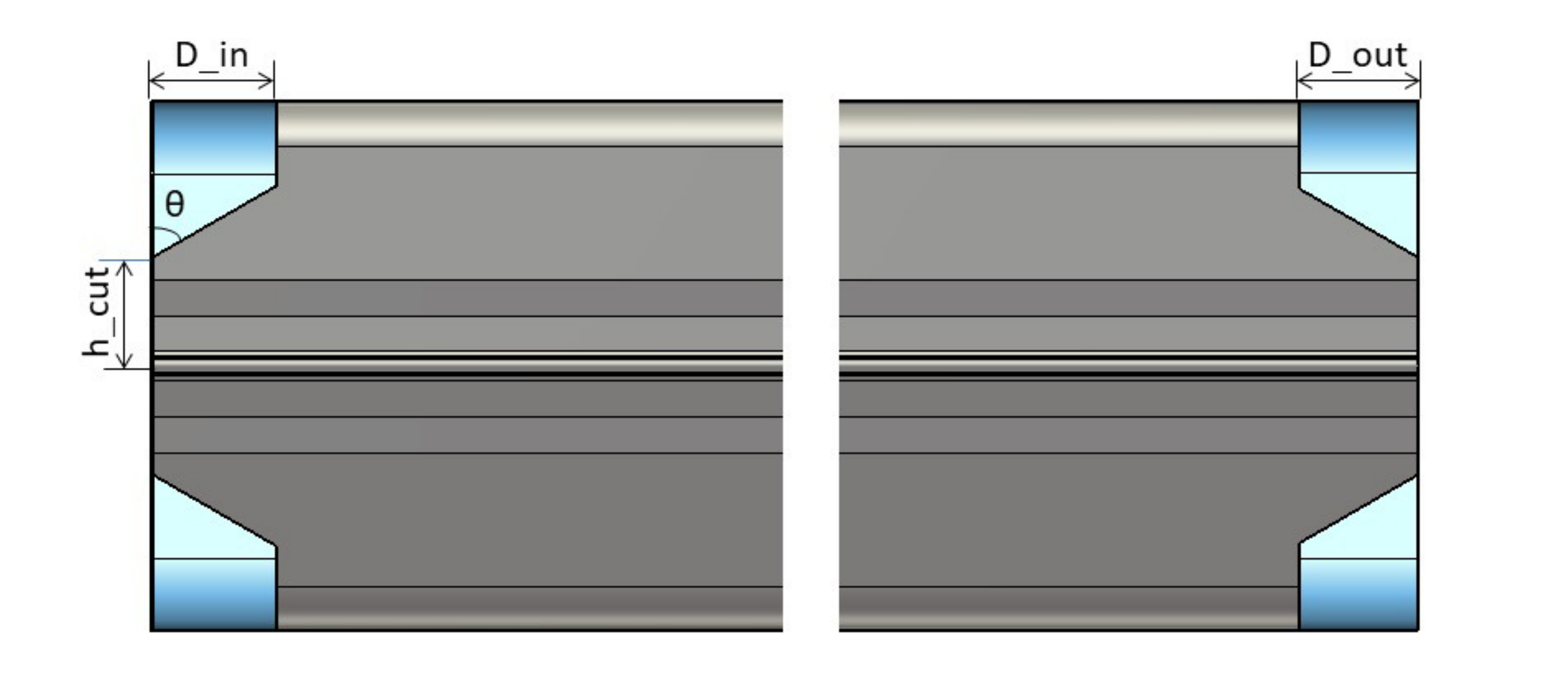}
		\caption{Design diagram of undercuts.}
		\label{The design diagram of undercut.}
	\end{figure}

	Another important task of an RF structure design is to tune the horizontal distribution of the electric field between electrodes along the longitudinal direction. The flat distribution of the electric field is extremely important for a frequency regulation during an operation. If the electric field is flat, when the depth of the tuner insertion or draw out is the same during the frequency regulation, the influence on the cavity voltage is also the same. The field distribution of the RFQ without undercuts is shown in Fig.~\ref{undercut12}. The electric field is low at both ends and high in the middle of the RFQ. The structure of undercuts is shown in Fig.~\ref{The design diagram of undercut.}, and Table~\ref{Parameters of tuners and undercuts.} lists the design parameters of the undercuts. The undercut design generally adopts a triangular type, which has the advantages of high mechanical strength and convenient processing. By cutting off the conductor part and increasing the volume of the vacuum part, as shown in Fig.~\ref{undercut12}, the electric field distribution is flat along the longitudinal direction after adding the undercuts.

	\subsection{Whole cavity simulation}\label{subsec:The whole cavity simulation}
	Thus far, the design of the cross-section and model of the RFQ cavity has been completed. To reduce the computing time, the electromagnetic structure design and electromagnetic simulation described in this paper are all carried out using the average aperture R0 instead of the vane-tip modulation data generated by RFQGen software in the dynamics simulation, the results of which are similar. To increase the accuracy of the results and make the model more similar to a real RFQ, the electrodes were chamfered, and the modulations of RFQ were added to the model in the following whole-cavity simulation.
	
	The RFQ cavity has 64 tuners, with a 3162.62-mm vane length and 3176.75-mm total length of the cavity and no frequency separation structure. The radial matching gap at the entrance of the RFQ (8.68 mm) and the fringe-field gap at the exit of the RFQ (5.45 mm) were added to the cavity. Both were determined using RFQGen software. CST software was used to simulate the entire cavity, and the Kp factor was found to be 1.69. The specific high-frequency parameters are listed in Table~\ref{The RF parameters of whole cavity RFQ.}.  
	
	It is necessary to determine whether the dipole mode affects the normal operation mode of the RFQ because it will be excited if its frequency is close to the operation frequency.
	In the final whole-cavity RF simulation, as shown in Table~\ref{The RF parameters of whole cavity RFQ.}, the minimum frequency interval $\Delta f$ between the dipole mode and working quadrupole mode is 2.144 MHz. The Q value of the cavity was 14,361. According to references ~\cite{ma2017design,microwave_engineering}, the effect of the dipole mode on the working quadrupole mode can be calculated through the following: 
	\begin{equation}
	\alpha = 1/\sqrt{1+{(Q\frac{2\Delta f}{f_0})}^2}
	\end{equation}
where $f_0$ represents the operating frequency of the cavity, which was 200 MHz in this case. Thus, the value of $\alpha$ is 0.32\%, which means that the nearest dipole mode has little effect on the operation mode. Therefore, there is no need for a mode separation structure.
	
	\begin{table}[htbp]
		\centering 
		\caption{The RF parameters of whole-cavity RFQ.} 
		\label{The RF parameters of whole cavity RFQ.}
			\begin{tabular}{lll} 
			\toprule
				Parameter         & Unit       & Value   \\
				\midrule
				H                 & mm         & 147.82  \\
				Frequency separation         & MHz        & 2.144 \\
				Q                 &        & 14361    \\ 
				Power loss        & kW         & 87.545   \\
				\bottomrule
			\end{tabular} 
	\end{table}

	\section{Conclusions}\label{sec:Conclusions}
	This paper takes the method of using RFQ to accelerate proton beam bombarding on lithium target to produce neutrons. The RFQ can be used as the front-end accelerator of AB-BNCT neutron source facility. The beam dynamics simulation of the RFQ was carried out using RFQGen software. At an operating frequency of 200 MHz, the transmission efficiency of a 20-mA proton beam accelerated from 30 keV to 2.5 MeV reached 98.7\%, and the vane voltage and maximum surface field were controlled within a reasonable range. By studying the margins of the parameters of the accelerator, it is shown that the transmission efficiency of the RFQ can be maintained at above 95\% under the condition that the parameters do not significantly change. For the RF structure design of the RFQ, the quadrilateral structure of the four-vane cavity was selected, which has the advantages of high mechanical strength, easy processing, simple water-cooling structure, and suitability for accelerating CW beams. In addition, to achieve the frequency regulation and field flatness regulation, tuners and undercuts were designed. Finally, the entire cavity with modulations and chamfers was simulated to meet the design requirements.
	
	This paper mainly describes the design of the main body of the RFQ but does not consider the connection between the RFQ and front-end or back-end facility. For the RFQ design, it is also necessary to add the design of the coupler structure and perform a multi-physical field coupling analysis~\cite{multiphysics} and a  secondary electron multiplication effect simulation.


\begin{thebibliography} {99}
	

	
	
	\bibitem{locher1936biological}	G. L. Locher, Biological effects and therapeutic possibilities of neutron. American Journal of Roentgenology {\bf 36}, 1 (1936). 
	
	\bibitem{Rogus1994Mixed}	R. D. Rogus, O. K. Harling, J. C. Yanch et al., Mixed field dosimetry of epithermal neutron beams for boron neutron capture therapy at the mitr-ii research reactor. Medical Physics {\bf 21}, 1611–1625 (1994)
	\href{https://doi.org/10.1118/1.597267}{doi: 10.1118/1.597267}
	
	\bibitem{KASESAZ2014132}	Yaser Kasesaz, Hossein Khalafi, Faezeh Rahmani et al.,	A feasibility study of the Tehran research reactor as a neutron source for BNCT. Applied Radiation and Isotopes {\bf 90}, 132–137 (2014).
	\href{https://doi.org/10.1016/j.apradiso.2014.03.028}{doi: 10.1016/j.apradiso.2014.03.028}
	
	\bibitem{1969A}	I. M. Kapchinskii, V. A. Teplyakov, A Linear Ion Accelerator with Spatially Uniform Hard Focusing. 1-17 (1969).
	
	\bibitem{2002Conceptual}	T. P. Wangler, J. E. Stovall, T. S. Bhatia et al., Conceptual design of an RFQ accelerator-based neutron source for boron neutron-capture therapy. Proceedings of the 1989 IEEE Particle Accelerator Conference {\bf 1}, 678–680 (2002).
	\href{https://doi.org/10.1109/PAC.1989.73220}{doi: 10.1109/PAC.1989.73220}
	
	\bibitem{2004TRASCO}	P. A. Posocco, M. Comunian, E. Fagotti, A. Pisent, TRASCO-RFQ as Injector for the SPES-1 Project. Proceedings of Linac–66-68 (2004).
	 
	\bibitem{2014Japan}	M. Yoshioka, T. Kurihara, S. Kurokawa et al.,
	Construction of the accelerator-based BNCT facility at the IBARAKI Neutron Medical Research Center Proceedings of LINAC2014, 230–232 (2014).
	
	\bibitem{IAEA}	IAEA, Current status of neutron capture therapy, (2001). \url{http://www-pub.iaea.org/MTCD/publications/PDF/te_1223_prn.pdf}
	
	\bibitem{aJEON2020633}	Byoungil Jeon, Jongyul Kim, Eunjoong Lee et al., Target-Moderator-Reflector system for 10-30 MeV proton accelerator-driven compact thermal neutron source: Conceptual design and neutronic characterization. Nuclear Engineering and Technology {\bf 52}, 633–646 (2020).
	\href{https://doi.org/10.1016/j.net.2019.08.019}{doi: 10.1016/j.net.2019.08.019}
	
	\bibitem{ZHU201857}	X. W. Zhu, H. Wang, Y. R. Lu et al., 2.5 MeV CW 4-vane RFQ accelerator design for BNCT applications. Nuclear Inst. and Methods in Physics Research, A {\bf 883}, 57–74 (2018).
	\href{https://doi.org/10.1016/j.nima.2017.11.042}{doi: 10.1016/j.nima.2017.11.042}
	
	
	\bibitem{1997Simulations}  L. M. Young, Simulations of the LEDA RFQ 6.7-MeV accelerator. Particle Accelerator Conference, 2752–2754 (1997).
	
	\bibitem{Comunian2008THE}  M. Comunian, A. Pisent, Legnaro et al., The IFMIF-EVEDA RFQ: Beam Dynamics Design. Proceedings of Linac (2008)
	
	\bibitem{2015DESIGN} N. K. Bultman, E. Pozdeyev, G.Morgan et al., Design of the FRIB RFQ. Ipac. (2015).
	
	\bibitem{RFQCodes} K. R. Crandall, T. P. Wangler, L. M. Young et al., RFQ Design Codes. Los Alamos National Laboratory. 2005.
	
	\bibitem{2007The}	R. H. Stokes, T. P. Wangler, K. R. Crandall, The Radio-Frequency Quadrupole - A New Linear Accelerator. IEEE Transactions on Nuclear Science {\bf 28} 1999–2003 (2007)
	\href{https://doi.org/10.1109/TNS.1981.4331575}{doi: 10.1109/TNS.1981.4331575}
	
	\bibitem{2012DESIGN}	Z. L. Zhang, H. W. Zhao, J. Wang et al., Design of a four-vane RFQ for China ADS project. Proceedings of Linac (2012)

	\bibitem{2015Japan}		J. Wei, H. Ao, N. Bultman et al., FRIB accelerator: Design and construction status. Proceedings of HIAT2015 (2015)

	\bibitem{aSNSRFQ}	A. Ratti, R. DiGennaro, L. Doolittle et al., Fabrication and testing of the first module of the SNS RFQ. International Linac Conference (2000)
	
	\bibitem{LI201738}	C. X. Li, Y. He, X. B. Xu et al., RF structure design of the China Material Irradiation Facility RFQ. Nuclear Inst. and Methods in Physics Research, A {\bf 869}, 38–45 (2017).
	\href{https://doi.org/10.1016/j.nima.2017.06.045}{doi: 10.1016/j.nima.2017.06.045}
	
	\bibitem{CHAKRABARTI2004599}	Alok Chakrabarti, Vaishali Naik, O. Kamigaito et al., The design of a four-rod RFQ LINAC for VEC-RIB facility. Nuclear Inst. and Methods in Physics Research, A {\bf 535}, 599–605 (2004).
	\href{https://doi.org/10.1016/j.nima.2004.06.158}{doi: 10.1016/j.nima.2004.06.158}
	
	\bibitem{2011Simulation}	Q. F. Zhou, Z. Kun, Y. R. Lu et al., Simulation and experiments of rf tuning of a 201.5 MHz four-rod RFQ cavity. Chinese Physics C {\bf 35}, 1042–1046 (2011).
	\href{https://doi.org/10.1088/1674-1137/35/11/012}{doi: 10.1088/1674-1137/35/11/012}

	\bibitem{IHRFQ} Z. S. Li., X. J. Yin, H. Du et al., The IH-RFQ for HIRFL-CSR injector. Nuclear Science and Techniques {\bf 29} 89 (2018) 
	\href{https://doi.org/10.1007/s41365-018-0416-y}{doi: https://doi.org/10.1007/s41365-018-0416-y}
	
	\bibitem{ma2017design}	W. Ma, L. Lu, X. B. Xu et al., Design of an 81.25 MHz continuous-wave radio-frequency quadrupole accelerator for Low Energy Accelerator Facility. Nuclear Inst. and Methods in Physics Research, A {\bf 847}, 130–135 (2017).
	\href{https://doi.org/10.1016/j.nima.2016.11.056}{doi: 10.1016/j.nima.2016.11.056}

	\bibitem{CST} CST Studio Suite.
	\url{http://www.cst.com}
	
\bibitem{microwave_engineering}	David M. Pozar, Microwave Engineering. John Wiley and Sons, Inc. (2011)
	
	\bibitem{multiphysics}	B. Zhao, B. zhang, S. P. Chen, et al., Multi-physics field coupling analysis of radio frequency quadrupole cavity. Nuclear Science and Techniques {\bf 43}, 30202 (2020)






\end{thebibliography}
\end{document}